# Synthesis and crystal growth of $Cs_{0.8}(FeSe_{0.98})_2$: a new iron-based superconductor with $T_c$=27K

A Krzton-Maziopa[1], Z Shermadini[2], E Pomjakushina[1], V Pomjakushin[3], M Bendele[2,4], A Amato[2], R Khasanov[2], H Luetkens[2], and K Conder[1]

[1]*Laboratory for Developments and Methods, Paul Scherrer Institute, CH-5232 Villigen PSI, Switzerland*

[2]*Laboratory for Muon Spin Spectroscopy, Paul Scherrer Institute, CH-5232 Villigen PSI, Switzerland*

[3]*Laboratory for Neutron Scattering, Paul Scherrer Institute, CH-5232 Villigen PSI, Switzerland*

[4]*Physik-Institut der Universität Zürich, Winterthurerstrasse 190, CH-8057 Zürich, Switzerland*

E-mail: kazimierz.conder@psi.ch

**Abstract.** We report on the synthesis of large single crystals of a new FeSe-layer superconductor $Cs_{0.8}(FeSe_{0.98})_2$. X-ray powder diffraction, neutron powder-diffraction and magnetization measurements have been used to compare the crystal structure and the magnetic properties of $Cs_{0.8}(FeSe_{0.98})_2$ with those of the recently discovered potassium intercalated system $K_xFe_2Se_2$. The new compound $Cs_{0.8}(FeSe_{0.98})_2$ shows a slightly lower superconducting transition temperature ($T_c$=27.4 K) in comparison to 29.5 in $K_{0.8}(FeSe_{0.98})_2$). The volume of the crystal unit cell increases by replacing K by Cs - the c-parameter grows from 14.1353(13) Å to 15.2846(11) Å. For the so far known alkali metal intercalated layered compounds ($K_{0.8}Fe_2Se_2$ and $Cs_{0.8}(FeSe_{0.98})_2$) the $T_c$ dependence on the anion height (distance between Fe-layers and Se-layers) was found to be analogous to those reported for As-containing Fe-superconductors and $Fe(Se_{1-x}Ch_x)$, where *Ch*=Te, S.

## 1. Inroduction

The recent discovery of the Fe-based superconductors has triggered a remarkable renewed interest for possible new routes leading to high-temperature superconductivity. As observed in the cuprates, the iron-based superconductors exhibit interplay between magnetism and superconductivity suggesting the possible occurrence of unconventional superconducting states. Other common properties are the layered structure and the low carrier density. Among the iron-based superconductors $FeSe_{1-x}$ has the simplest structure with layers in which Fe cations are tetrahedrally coordinated by Se [1]. The superconducting transition temperature ($T_c$) of 8 K was found to dramatically increase under pressure reaching a maximum of 37K, with a rate of $dT_c/dP$~9.1 K/GPa -- the highest among all the Fe-based superconductors [2]. Additionally, it was found an applied pressure modifies the electronic phase diagram of $FeSe_{1-x}$ and induces static magnetic order which can coexist with superconductivity [3]. Moreover, the substitution of Te for Se leads to an increase of $T_c$ up to 14K [4]. Since Te is larger than Se, this effect could not be considered as corresponding to an externally applied pressure. It was also found that substitution of S for Se increases slightly $T_c$ [5]. Comprehensive studies of substitutions on iron site in $M_xFe_{1-x}Se_{0.85}$ with the non-transition metals M=Al, Ga, In, Sm, Ba and transition metals M=Ti, V, Cr, Mn, Co, Ni and Cu, were performed by Wu et al [6]. In the first case it was stated that the observed slight change of $T_c$ (within ±2K) depends on the size of the substituting ions and the doping level suggesting the importance of the lattice deformation on the superconducting properties. All the systems with transition metals (with exception of Mn) do not exhibit superconductivity. In the case of M=Cu already 1.5% of doping suppress superconductivity [7] and with 10% the compound becomes a Mott insulator [8]. Transition-metal doping was also reported in the case of $Fe_{1-x}Se_{0.5}Te_{0.5}$. Doping with Co (0.05≤x≤0.2) and Ni (0.05≤x≤0.1) both suppress $T_c$ and lead to a metal-insulator transition [9]. Doping with Cu (x=0.05) destroys superconductivity whereas magnetic Mn (x=0.05) slightly increases $T_c$ [10]. The alkali metal doped $Na_{0.1}FeSe$, where the Na-ions are intercalated between FeSe layers, was found to be superconducting with Tc =8.3 K [11].

Very recently superconductivity at above 30K was found in $K_{0.8}Fe_2Se_2$ [12]. This is so far the highest $T_c$ for Fe–chalcogenides, even though the superconducting fraction is low and the transition is broad. The potassium-ions are intercalated between FeSe layers increasing the distance between them. The Fe-Fe layer distance is 7.0184 Å in comparison with 2.76.17 Å in FeSe [13]. The intercalation of K also increases the Fe-Se bond length within the layers by 2.15%. It is reported [12] that single crystals of several $mm^3$ could be grown from the self-flux.

In the present work we report on the synthesis and crystal growth of a new analog compound with Cs intercalated between FeSe layers. In comparison with the work of Guo et al. [12] we managed by this substitution to significantly increase the superconducting fraction by only slightly diminishing of the critical temperature.

## 2. Experimental details

Single crystals of both potassium and cesium intercalated iron selenides of nominal compositions $Cs_{0.8}(FeSe_{0.98})_2$ and $K_{0.8}(FeSe_{0.98})_2$ were grown from the melt using the Bridgeman method. Ceramic rods of iron selenide starting material were prepared by solid state reaction technique [13]. The nominal stoichiometry of the starting material that is $FeSe_{0.98}$ was chosen based in view of our previous studies [13] which demonstrated that for this particular Fe/Se ratio the content of secondary phases is the smallest. High purity (at least 99.99%, Alfa) powders of iron and selenium were mixed in an 8g batch, pressed into rods, sealed in evacuated quartz ampoules and annealed at 700°C during 15h.

The initially treated material was then grounded in an inert atmosphere, pressed again into rods, sealed in evacuated quartz ampoules and thermally treated at 700°C over 48h followed by further annealing at 400°C for another 36h.

For the single crystal synthesis a piece of the ceramic rod of $FeSe_{0.98}$ was sealed in double walls evacuated silica ampoule with the pure alkali metals (either potassium or cesium of at least 99.9% purity, Chempur). The quantity of alkali metal used for the synthesis depended on the desired stoichiometry of the final compound. The ampoules were annealed at 1030°C over 2h for homogenization. Afterwards the melt was cooled down to 750°C with the rate of 6°C/h and then cooled down to room temperature with the a rate 200°C/h. Well formed black crystal rods of 7 mm diameter (diameter of the quartz ampoules) were obtained which could be easily cleaved into plates with flat shiny surfaces.

The $Cs_{0.8}(FeSe_{0.98})_2$ and $K_{0.8}(FeSe_{0.98})_2$ crystals were characterized by powder x-ray diffraction (XRD) using a D8 Advance Bruker AXS diffractometer with Cu $K_\alpha$ radiation. For these measurements a fraction of each crystal was cleaved, powderized, and loaded into the low background airtight specimen holder in a He-glove box to protect the powder from oxidation. The $K_{0.8}(FeSe_{0.98})_2$ polycrystalline sample, which was synthesized the same way as proposed by Guo et al. [12], was additionally studied by means of neutron powder diffraction (NPD) at the SINQ spallation source of the Paul Scherrer Institute (PSI, Switzerland) using the high-resolution diffractometer for thermal neutrons HRPT [14] with the neutron wavelengths λ=1.494 and 1.886Å. The sample was loaded into the vanadium container with indium sealing in the He-glove box. The refinements of the crystal structure parameters were done using the FULLPROF program [15] with the use of its internal tables.

The superconducting transition has been detected by AC susceptibility by using a conventional susceptometer. The sample holder contains a standard coil system with a primary excitation coil (1300 windings, 40mm long) and two counter-winded pick-up coils (reference and sample coil, each 10mm long and 430 windings) which are connected to a lock-in amplifier. The used frequency was 144 Hz and the sample holder diameter was 5 mm. Measurements were performed by heating the sample at a rate of 9 K/h. The susceptometer was calibrated using the superconducting transition of a lead sample showing a 100% superconducting fraction. The raw data then were normalized to the sample volume relative to the one of the Pb calibration specimen.

## 3. Results and Discussion

Room-temperature XRD experiments revealed, that the crystals do not contain any impurity phases. The only detected phase is the tetragonal phase of $ThCr_2Si_2$ type (space group *I/4mmm*). The results of the Rietveld refinement of the XRD patterns are shown in figure 1 and figure 2 for $Cs_{0.8}(FeSe_{0.98})_2$ and

K$_{0.8}$(FeSe$_{0.98}$)$_2$, correspondingly. A broad halo on XRD pattern around 20 deg is caused by a sample holder with a plastic dome. This area was excluded from the refinements. For the refinement it was assumed, that all Fe and Se sites are fully occupied. The crystallographic data for Cs$_{0.8}$(FeSe$_{0.98}$)$_2$, K$_{0.8}$(FeSe$_{0.98}$)$_2$ crystals and K$_{0.8}$(FeSe$_{0.98}$)$_2$ polycrystalline are summarized in the table 1.

**Table 1.** Structural parameters for Cs$_{0.8}$(FeSe$_{0.98}$)$_2$, K$_{0.8}$(FeSe$_{0.98}$)$_2$ powderized crystals and K$_{0.8}$(FeSe$_{0.98}$)$_2$ powder at 290 K obtained from XRD and NPD, correspondingly. Space group *I4/mmm* (no. 139), Fe in (*4d*) position (0, 0.5, 0.25); Se in (*4e*) position (0, 0, z), Cs/K in (*2a*) position (0, 0, 0). The atomic displacement parameters for all atoms were constrained to be the same.

| | Cs$_{0.8}$(FeSe$_{0.98}$)$_2$ (XRD) powderized crystal | K$_{0.8}$(FeSe$_{0.98}$)$_2$ (XRD) powderized crystal | K$_{0.8}$(FeSe$_{0.98}$)$_2$ (NPD) polycrystalline |
|---|---|---|---|
| a (Å) | 3.9601(2) | 3.9092(2) | 3.9038(1) |
| c (Å) | 15.2846(11) | 14.1353(13) | 14.1148(6) |
| Se, z | 0.3439(3) | 0.3503(3) | 0.3560(3) |
| Cs/K occupancy | 0.771(7) | 0.792(10) | 0.737(20) |
| B (Å$^2$) | 3.37(9) | 3.16(9) | 1.63(4) |
| R$_p$, R$_{wp}$, R$_{exp}$ | 4.35, 6.00, 3.09 | 3.84, 5.10, 2.99 | 4.91, 6.71, 3.25 |
| $\chi^2$ | 3.77 | 2.90 | 4.26 |

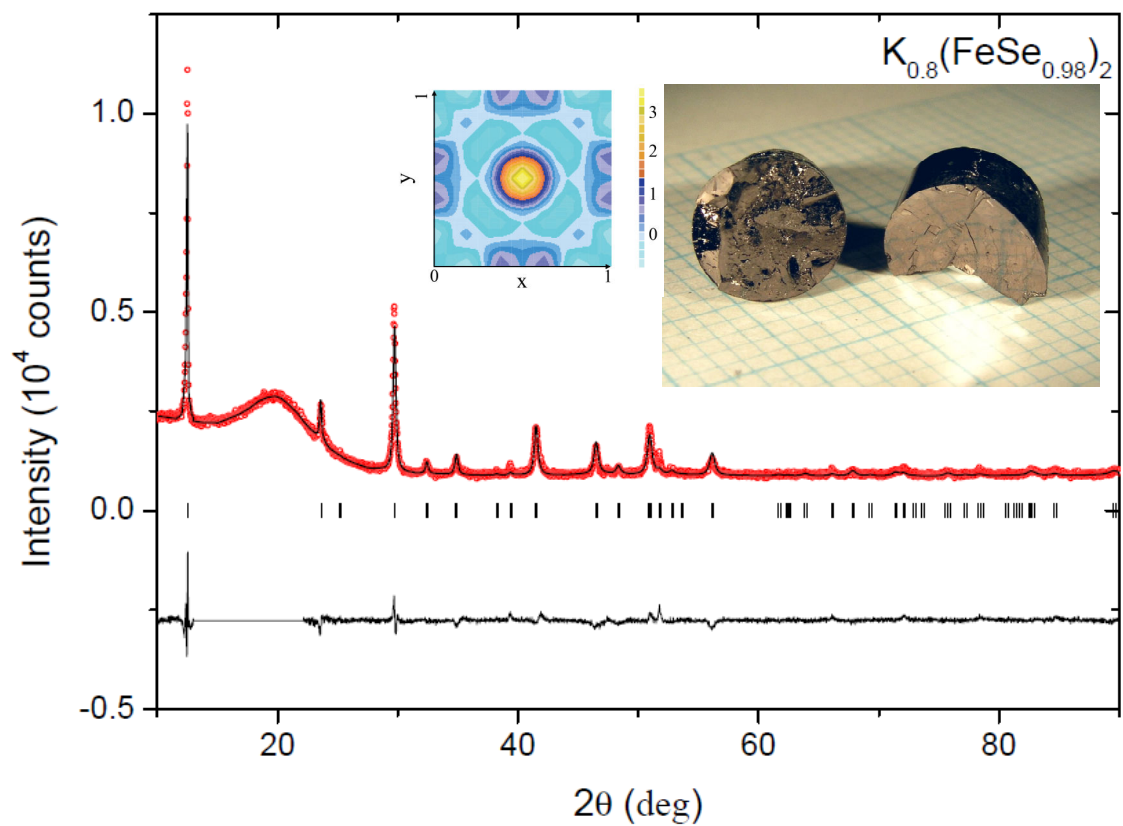

**Figure 1.** (Color online) Rietveld refinement pattern (upper–red) and difference plot (lower–black) of the X-ray diffraction data for the crystal with the nominal composition of K$_{0.8}$(FeSe$_{0.98}$)$_2$. The rows of ticks show the Bragg peak positions for the *I4/mmm* phase. The left insert shows the difference Fourier density map at z=1/2 slice obtained from NPD data showing the presence of K at (0.5, 0.5, 0.5),

colored scale shows scattering density in fm. The right insert shows a picture of the cleaved $K_{0.8}(FeSe_{0.98})_2$ crystal.

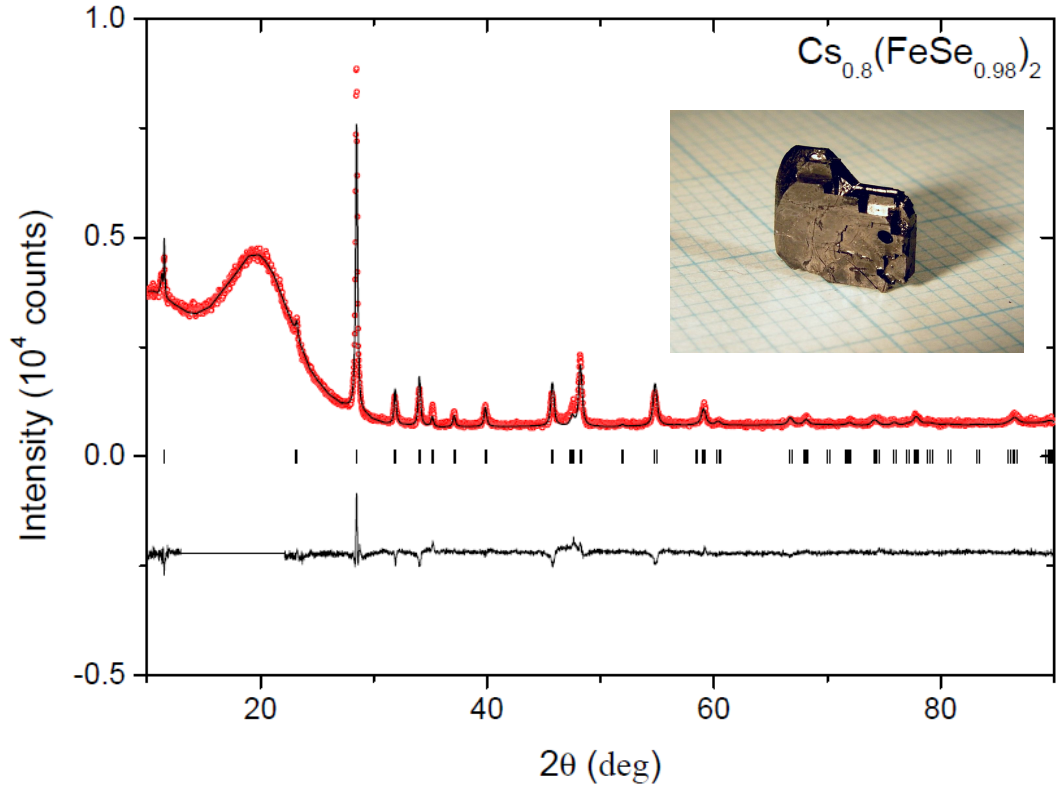

**Figure 2**. (Color online) Rietveld refinement pattern (upper–red) and difference plot (lower–black) of the X-ray diffraction data for the crystal with the nominal composition of $Cs_{0.8}(FeSe_{0.98})_2$. The rows of ticks show the Bragg peak positions for the *I4/mmm* phase. The insert shows a picture of the a piece of $Cs_{0.8}(FeSe_{0.98})_2$ crystal.

One can note that the atomic displacement parameters (ADP) refined from XRD are quite large in comparison with the ones refined from NPD. We believe that this results from a slight degradation of the samples during the XRD measurements due to the non-ideal sealing of the plastic container, inasmuch as the samples are extremely air sensitive. It is also supported by the fact that both samples show a pronounced strain-like diffraction peak broadening. The enhanced ADP can lead to systematic errors in determination of the site occupancies. By substituting K to Cs the lattice expands predominately in the c-direction, due to the larger ionic radius of Cs (1.78 Å) compared to K (1.51Å).

Figure 3 shows the temperature dependence of the AC susceptibility ($\chi'$) for single crystals of $K_{0.8}(FeSe_{0.98})_2$ and $Cs_{0.8}(FeSe_{0.98})_2$. The onset of the critical temperature has been determined to $T_{c,onset}$ = 27.4 K and 29.5 K for the Cs and K intercalated compounds, respectively. It has to be noticed that the diamagnetic signal is larger in the case of the Cs intercalated FeSe which might point to a bigger

Meissner fraction. A further investigation if this effect is intrinsic to the $K_{1-x}(FeSe_{0.98})_2$ and $Cs_{1-x}(FeSe_{0.98})_2$ families or if the superconducting fraction sensitively depends on the composition x is underway.

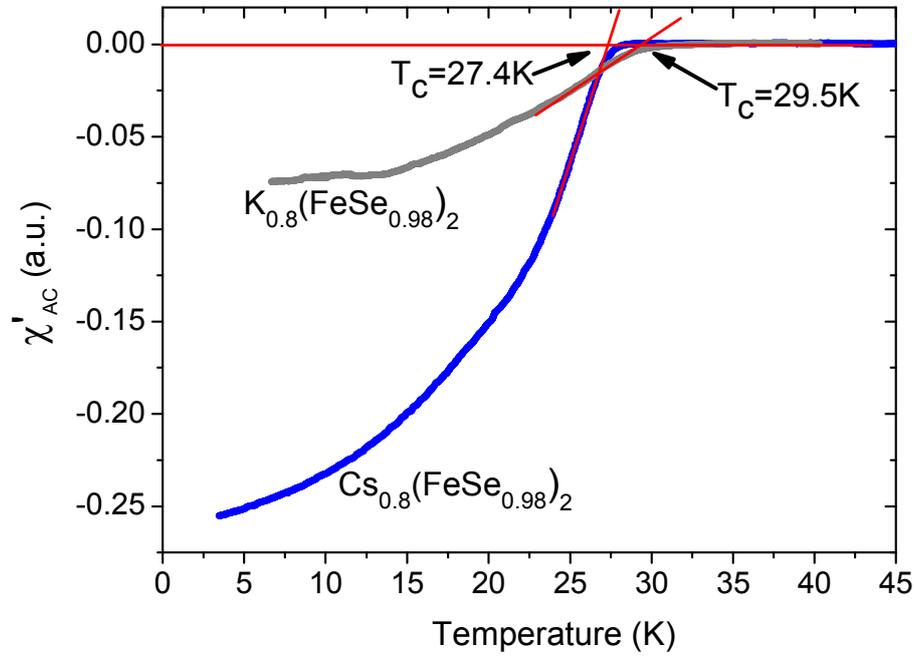

**Figure 3.** (Color online) Temperature dependence of the AC susceptibility ($\chi'$) for single crystalline $K_{0.8}(FeSe_{0.98})_2$ and $Cs_{0.8}(FeSe_{0.98})_2$. The signal has been normalized to a superconducting Pb specimen as described in the text.

Intercalation of the alkali metals into the FeSe causes serious structural changes. It was proved by Mizuguchi et al. [16] that in iron pnictides and chalcogenides the critical temperature can be correlated with so known "anion height" which is the distance between Fe and chalcogenide (or pnictide) layers in the structure. Figure 4 presents a line (taken from Ref. [16]) being a best fit to the experimental data obtained for over 15 differed Fe-based superconducting compounds. The curve shows a relatively sharp peak around 1.38Å with a maximum transition temperature $T_c \approx 55K$ (for $NdFeAsO_{0.83}$). The open symbols in figure 4 depict the anion height to $T_c$ correlation of the samples synthesized in this work and those presented in our previous reports [17,18]. Apparently the newly synthesized compounds with intercalated Cs and K follow very well the universal trend. The tendency in the series $FeSe_{0.98}$-$Cs_{0.8}(FeSe_{0.98})_2$-$K_{0.8}(FeSe_{0.98})_2$ is analogous to those reported for high pressure measurements [2] (filled circles in figure 4). The steep slope of $T_c$ as a function of anion height suggests that even much higher superconducting transition temperatures might be found in the newly discovered FeSe based systems by applying either chemical (substitutional) or hydrostatic pressures. Another not yet explored aspect is the relation of magnetism and superconductivity in this system.

Anyhow, whether or not superconductivity appears in this system in closed proximity to an antiferromagnetically ordered state like in other Fe-based superconductors awaits further investigation. Fortunately, the synthesis method described here is able to provide large single crystals which will allow to answer the above risen questions in detail by applying bulk methods like neutron scattering or muon spin rotation both at ambient and high pressures.

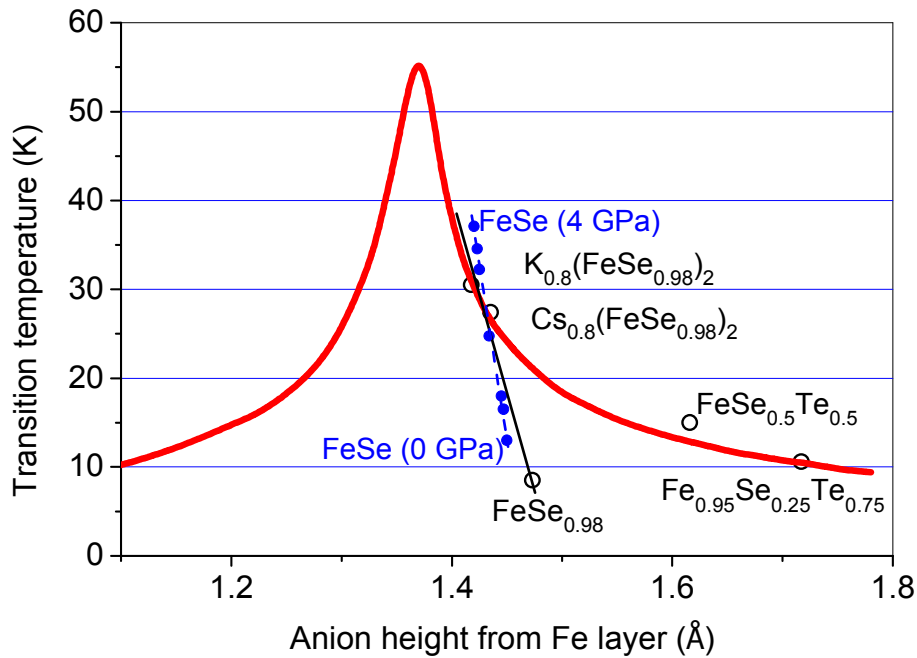

**Figure 4**. (Color online) Dependence of $T_c$ on the distance between Fe and the chalcogenide/ pnictide layers. The red line shows the dependence as presented by Mizuguchi et al. [16] for typical Fe-based superconductors. The filled circles indicate the data obtained for FeSe under high pressure [2]. The open points indicate the data obtained by this work and our previous reports [17, 18].

## 4. Summary

In conclusion, a new Cs intercalated iron selenide superconductor ($Cs_{0.8}(FeSe_{0.98})_2$) was synthesized by the Bridgeman method in form of large single crystals. The $Cs_{0.8}(FeSe_{0.98})_2$ compound represents the second member of the alkali metal chalcogenide family. In comparison with the K-analog a larger lattice volume is observed and $T_{c,onset}$ = 27.4 K is only slightly decreased compared to $K_{0.8}Fe_2Se_2$. The large high quality crystals obtained by the method described will allow an in-detail study of fundamental magnetic and superconducting properties in this new FeSe family.

**Acknowledgements**

The authors thank the Sciex-NMS[ch] (Project Code 10.048) and NCCR MaNEP for the support of this study. This study was partly performed at Swiss neutron spallation SINQ of Paul Scherrer Institute PSI (Villigen, PSI). We acknowledge the allocation of the beam time at the HRPT diffractometer of the Laboratory for Neutron Scattering (PSI, Switzerland).